\documentclass[utf8]{frontiersSCNS}
\usepackage{url,lineno,microtype,subcaption}
\usepackage[onehalfspacing]{setspace}
\usepackage[inline]{trackchanges}

\def\firstAuthorLast{Oliveira {et~al.}} %use et al only if is more than 1 author
\def\Authors{Denny M. Oliveira\,$^{1,2*}$, 
			 Eftyhia Zesta\,$^{2}$, 
			 Piyush M. Mehta\,$^{3}$,
			 Richard J. Licata\,$^{3}$,
			 Marcin D. Pilinski\,$^{4}$,
			 W. Kent Tobiska\,$^{5}$, and
			 Hisashi Hayakawa\,$^{6,7,8,9}$}

\def\Address{$^{1}$Goddard Planetary Heliophysics Institute, University of Maryland, Baltimore County, Baltimore, MD, United States, \\
			$^{2}$Geospace Physics Laboratory, NASA Goddard Space Flight Center, Greenbelt, MD, United States, \\
			$^{3}$West Virginia University, Morgantown, WV, United States, \\
			$^{4}$Laboratory for Atmospheric and Space Physics, University of Colorado, Boulder, CO, United States, \\
			$^{5}$Space Environment Technologies, Pacific Palisades, CA, United States, \\
			$^{6}$Institute for Space-Earth Environmental Research, Nagoya University, Nagoya, 4648601, Japan, \\
			$^{7}$Institute for Advanced Research, Nagoya University, Nagoya, 4648601, Japan, \\
			$^{8}$UK Solar System Data Centre, Space Physics and Operations Division, RAL Space, Science and Technology Facilities Council, Rutherford Appleton Laboratory, Harwell Oxford, Didcot, Oxfordshire, OX11 0QX, UK, \\
			$^{9}$Nishina Center, Riken, Wako, 3510198, Japan}

\def\corrAuthor{Denny Oliveira}
\def\corrEmail{denny.m.deoliveira@nasa.gov}

\usepackage[breaklinks, colorlinks = true,
            linkcolor = red,
            urlcolor  = red,
            citecolor = red,
            anchorcolor = white,
            breaklinks]{hyperref}

\begin{document}
	\onecolumn
	\firstpage{1}

	\title[Predicting thermosphere density during extreme magnetic storms]{The current state and future directions of modeling thermosphere density enhancements during extreme magnetic storms} 

	\author[\firstAuthorLast ]{\Authors} %This field will be automatically populated
	\address{} %This field will be automatically populated
	\correspondance{} %This field will be automatically populated

	\extraAuth{}

	\maketitle

	\begin{abstract}

		\section{}

		Satellites, crewed spacecraft and stations in low-Earth orbit (LEO) are very sensitive to atmospheric drag. A satellite’s lifetime and orbital tracking become increasingly inaccurate or uncertain during magnetic storms. Given the planned increase of government and private satellite presence in LEO, the need for accurate density predictions for collision avoidance and lifetime optimization, particularly during extreme events, has become an urgent matter and requires comprehensive international collaboration. Additionally, long-term solar activity models and historical data suggest that solar activity will significantly increase in the following years and decades. In this article, we briefly summarize the main achievements in the research of thermosphere response to extreme magnetic storms occurring particularly after the launching of many satellites with state-of-the-art accelerometers from which high-accuracy density can be determined. We find that the performance of an empirical model with data assimilation is higher than its performance without data assimilation during all extreme storm phases. We discuss how forecasting models can be improved by looking into two directions: first, to the past, by adapting historical extreme storm datasets for density predictions, and second, to the future, by facilitating the assimilation of large-scale thermosphere data sets that will be collected in future events. Therefore, this topic is relevant to the scientific community, government agencies that operate satellites, and the private sector with assets operating in LEO.

	\end{abstract}

	\section{Introduction}

		During magnetic storms, large amounts of magnetospheric energy enters the ionosphere-thermosphere system at high latitudes through field-aligned currents \citep{Prolss2011,Emmert2015}. Since ionospheric currents are intensified, the interaction of the moving plasma with the local neutral gas is enhanced through the collision between ions and neutral molecules, thereby further heating the neutral atmosphere. The atmosphere then expands upwards and satellites flying in higher regions experience increased atmospheric drag forces, which in turn intensify orbital drag effects. These effects ultimately decrease the satellite's life time and introduce significant errors in orbital tracking which increase as the storm becomes more intense \citep{Doornbos2006,Zesta2016b,Nwankwo2021}. \par

		The third satellite in the Sputnik series provided arguably the first observations of storm-induced orbital drag effects. \cite{Jacchia1959} observed a strong decay of Sputnik 1958$\delta$1 due to increased atmospheric drag forces. These pioneer observations led to the creation of many thermospheric empirical models throughout the decades, including the \cite{Jacchia1970} model and subsequent series; the Mass Spectrometer Incoherent Scatter model series developed by \cite{Hedin1987} and later improved by the Naval Research Laboratory to become the Mass Spectrometer Incoherent Scatter Extended model \citep[][]{Picone2002}, and the Drag Temperature Model developed by \cite{Bruinsma2015}. In this work, we will discuss results provided by the High Accuracy Satellite Drag Model \citep[HASDM;][]{Storz2005}, and the improved version of the Jacchia model series described by \cite{Bowman2008}, henceforth JB2008. \par

		The understanding of the thermosphere response to magnetic storms and the accurate capability of predicting subsequent satellite orbital drag effects is of paramount interest of, e.g., the U.S. Federal Government and the private sector \citep{NSWS2015,NSWAP2015,Cakaj2021}. The correct orbital tracking of low-Earth orbit (LEO) satellites particularly in a time window of 72 hours into the storm is of great interest for taking actions such as pre-determined maneuvers as a means to avoid a satellite's collision with debris in space or even with other satellites \citep{Pardini2009,Wang2010c,Lewis2019}. Such actions may be characterized as an important tool for preventing the occurrence of the Kessler Syndrome. First introduced by \cite{Kessler1978}, the Kessler Syndrome suggests that by the dawn of the 22nd century LEO regions will pose high risks for satellite traffic due to the high probability of satellite collisions with significantly increased space debris levels. As we will discuss later, the need for accurate predictions of satellite orbital track during extreme magnetic storms is twofold: (i), the number of satellites in LEO has been and will be significantly increased; and (ii), prediction models and historical data suggest solar activity will increase in the next years and decades. \par

		In this paper, we will focus on four major LEO spacecraft missions that carried/carry high-precision accelerometers that can be used to obtain high-quality density data. These satellites are named CHAllenge Mini-satellite Payload \citep[CHAMP;][]{Reigber2002a}, Gravity Recovery And Climate Experiment \citep[GRACE;][]{Tapley2004a}, Gravity field and steady-state Ocean Circulation Explorer \citep[GOCE;][]{Drinkwater2003}, and Swarm \citep{Siemes2016}. We will then pay particular attention to model performance during the only seven extreme magnetic storms observed by CHAMP and GRACE.

	\section{Data and models}

		\subsection{Data}

			CHAMP was launched in July 2000 at the initial altitude 456 km completing a full orbit around Earth in 90 minutes with orbit inclination 87.25$^\circ$. CHAMP completed a longitudinal cycle in $\sim$130 days. The acceleration measurement precision by CHAMP was 3.0$\times10^{-8}$ m/s$^2$ with 0.1 Hz cadence \citep{Bruinsma2004}. CHAMP re-entered in September 2010. \par

			The GRACE mission was composed by two twin satellites, named GRACE-A and -B, that were launched in March 2002. GRACE-B followed GRACE-A within a controlled distance of $\sim$220 km. Therefore, densities derived by both spacecraft were generally very similar. We then use data from the first spacecraft, hereafter termed GRACE data. The initial altitudes of GRACE were around 500 km. The orbital period of GRACE was 95 minutes with orbit declination 89.5$^\circ$. The longitudinal coverage by GRACE was usually completed within $\sim$160 days. The GRACE acceleration precision and cadence are 10 times higher in comparison to CHAMP \citep{Flury2008}. GRACE re-entered in March 2018. \par

			GOCE was launched in March 2009 at the initial altitude 270 km and orbit inclination 96.5$^\circ$. The GOCE accelerometer precision was as high as 1.0$\times10^{-11}$ m/s$^2$, with measurements being performed within the bandwith 0.005-0.1 Hz \citep{Bruinsma2014}. GOCE re-entered in October 2013. \par

			The Swarm mission is composed of three identical satellites that were launched after November 2013 at orbits within 480-538 km altitude with orbital inclination $\sim$88$^\circ$. The Swarm acceleration precision is 1.0$\times10^{-11}$ m/s$^2$ within a bandwith whose upper limit is 0.1 Hz \citep{Siemes2016}. As a simple choice, we use only data from Swarm A, hereafter Swarm. Swarm is the only mission analyzed in this study that is still in operation. \par

		\subsection{Models}

			The JB2008 empirical model computes thermospheric mass density using as inputs solar indices and the 3-hour time resolution ap index accounting for low geomagnetic activity. When geomagnetic activity intensifies (Dst $<$ --75 nT), the model switches over to Dst as the proxy for geomagnetic activity \citep{Bowman2008}. The solar indices used by JB2008 map the energy input into the thermosphere provided by different solar irradiance sources, which in turn affect density variability in the most important thermospheric layers \citep{Tobiska2008}. \par

			HASDM dynamically calibrates densities by assimilating the observed drag effects on a carefully selected set of low-perigee calibration satellites and it solves for the global thermospheric neutral density with 3-hour cadence by “correcting” the JB2008 density prediction. Typically, 75 calibration satellites are used to fit and correct the JB2008 density and the greater the number of calibration satellites, the higher the accuracy. Because HASDM is able to mitigate uncertainty in atmospheric effects, by resolving the differences between calibration satellites and the core JB2008 model at each epoch (time step), real-time nowcasting specification of the LEO drag environment has improved significantly \citep[e.g.,][]{Storz2005,Bowman2008,Licata2021b,Calabia2021}. \par

			Forecasting is achieved with the JB2008 model, driven by forecasted solar and geomagnetic drivers, with uncertainties growing significantly a few hours post epoch. \cite{Marcos2010} demonstrated with detailed comparisons and validation that JB2008 is the most accurate empirical model. In JB2008, the 1-sigma uncertainties at 400 km for a given epoch dropped from 15\% to 8\%. The use of JB2008 in HASDM, and the assimilation of calibration satellite data into the initial JB2008 density solutions, further dropped the comparable HASDM density uncertainties to less than 5\% for current epoch, lower altitude, and quiet atmosphere. \par

			Although HASDM is not available for direct use by the general scientific community, Space Environment Technologies (SET) has made global density outputs available to the public \citep{Tobiska2021}. The SET HASDM density data base, available from 2000 to 2019, covers altitudes in the range 175-825 km, with resolution of 15$^\circ$ (longitude), 10$^\circ$ (latitude), and 25 km (altitude) \citep{Tobiska2021,Licata2021a}.

	\section{Satellite coverage in LEO during magnetic storms, including extreme events}

\begin{figure}[]
			\centering
			\includegraphics[width=15cm]{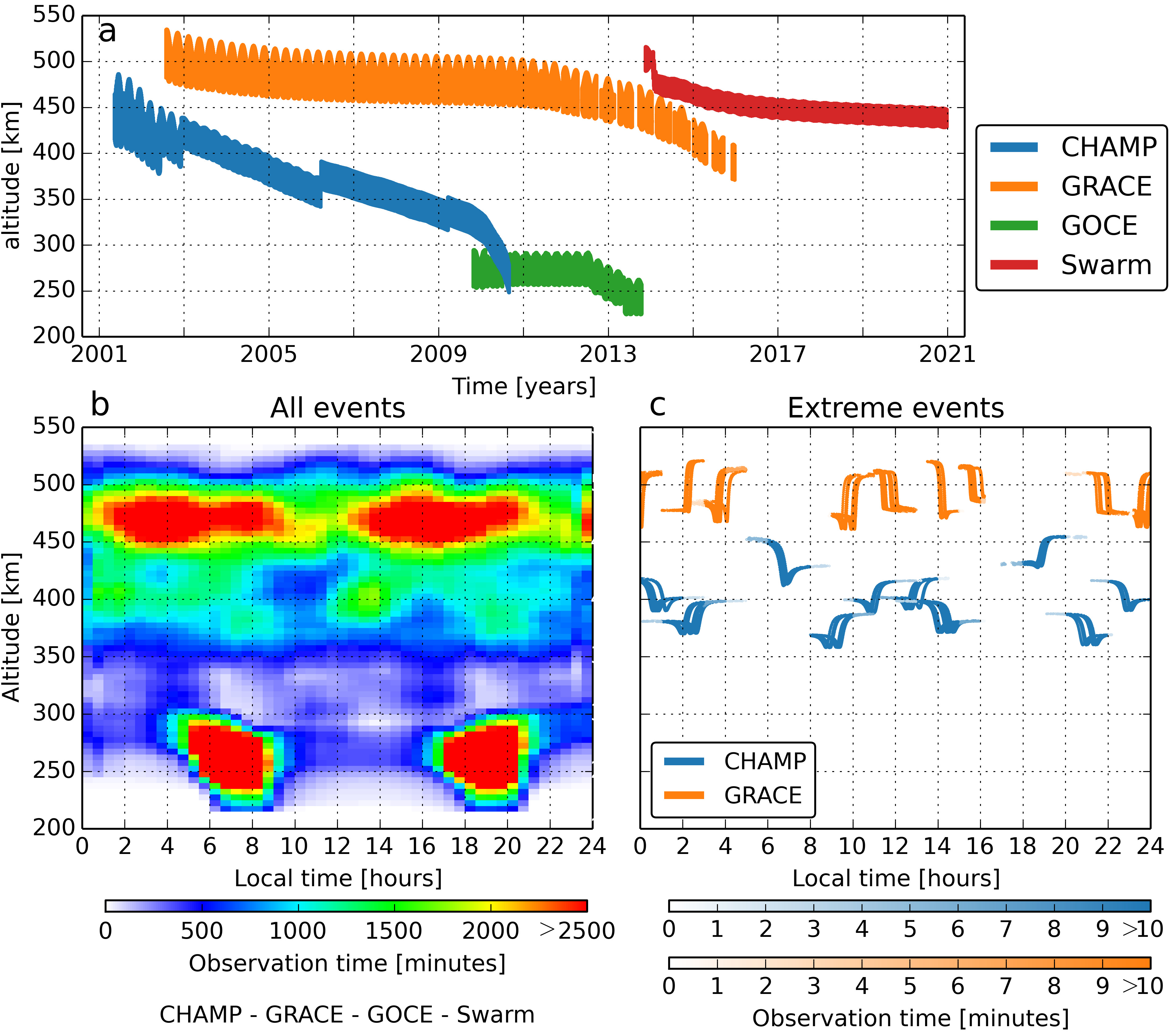}% This is a *.eps file
			\caption{Top panel: CHAMP, GRACE, GOCE, and Swarm altitudes. Bottom row, left panel: superposed epoch analysis of satellite observation times plotted as a function of local times and altitudes. The bin sizes are 5 km and 0.5 hour. Right panel: the same as plotted in the left panel, but for the 7 extreme magnetic storms (minimum SYM-H $\leq$ --250 nT) observed by CHAMP and GRACE.}
			\label{xht}
		\end{figure}

		Figure \ref{xht}a summarizes the altitudes plotted as a function of time of the four LEO satellites since the launching of CHAMP in May 2001 to December 2021. Although there has been significant coverage from $\sim$250 to $\sim$540 km altitude, this coverage is relatively sparse. As noted by \cite{Bruinsma2021}, the most well-covered storm was a mild event that occurred in April 2010, with CHAMP and GOCE around 300 km altitude, and GRACE around 480 km altitude, all with very limited local solar time coverage. \par

		Figure \ref{xht}b shows the observation time distributions as a function of altitude and local times for over 320 magnetic storms. The altitude data were collected within the 72-hour interval after the storm main phase onset as described by \cite{Oliveira2017c}. There is significant coverage above 400 km due to CHAMP, GRACE, and Swarm. Additionally, there is high coverage by GOCE between $\sim$180 km and $\sim$300 km concentrated around the dawn and dusk sectors. The coverage between $\sim$300 km and $350$ km is very low and comes from the late days of CHAMP's observations. Many advancements in our understanding of the thermosphere response to magnetic storms in the past two decades come from these observations, including, e.g., thermosphere heating in polar cusp regions \citep{Luhr2004a}, traveling atmospheric disturbance (TAD) propagation from high- to low-latitude regions \citep{Bruinsma2007,Fujiwara2006}, orbital drag effects induced by magnetic storms \citep{Krauss2015,Oliveira2019b}, high-latitude density enhancements due to magnetospheric compressions by solar wind pressure pulses \citep{Connor2016,Shi2017}, and thermosphere global time response to magnetic storms \citep{Sutton2005,Sutton2009a,Oliveira2017c}. \par

		Altitude coverages of all extreme events (minimum SYM-H $\leq$ --250 nT) ever observed by LEO spacecraft with high-accuracy accelerometers are shown in Figure \ref{xht}c. The figure shows observation time plotted as a function of local time and altitude for the extreme storms observed by CHAMP (7 events) and GRACE (6 events). In comparison with all events, the LEO observational coverage during extreme magnetic storms is remarkably sparser and briefer, particularly due to the rarity of extreme events and the low number of concurring LEO missions operating during these events.

	\section{Comparing model performance during the extreme events observed by CHAMP and GRACE}

\begin{figure}[]
			\centering
			\includegraphics[width=15cm]{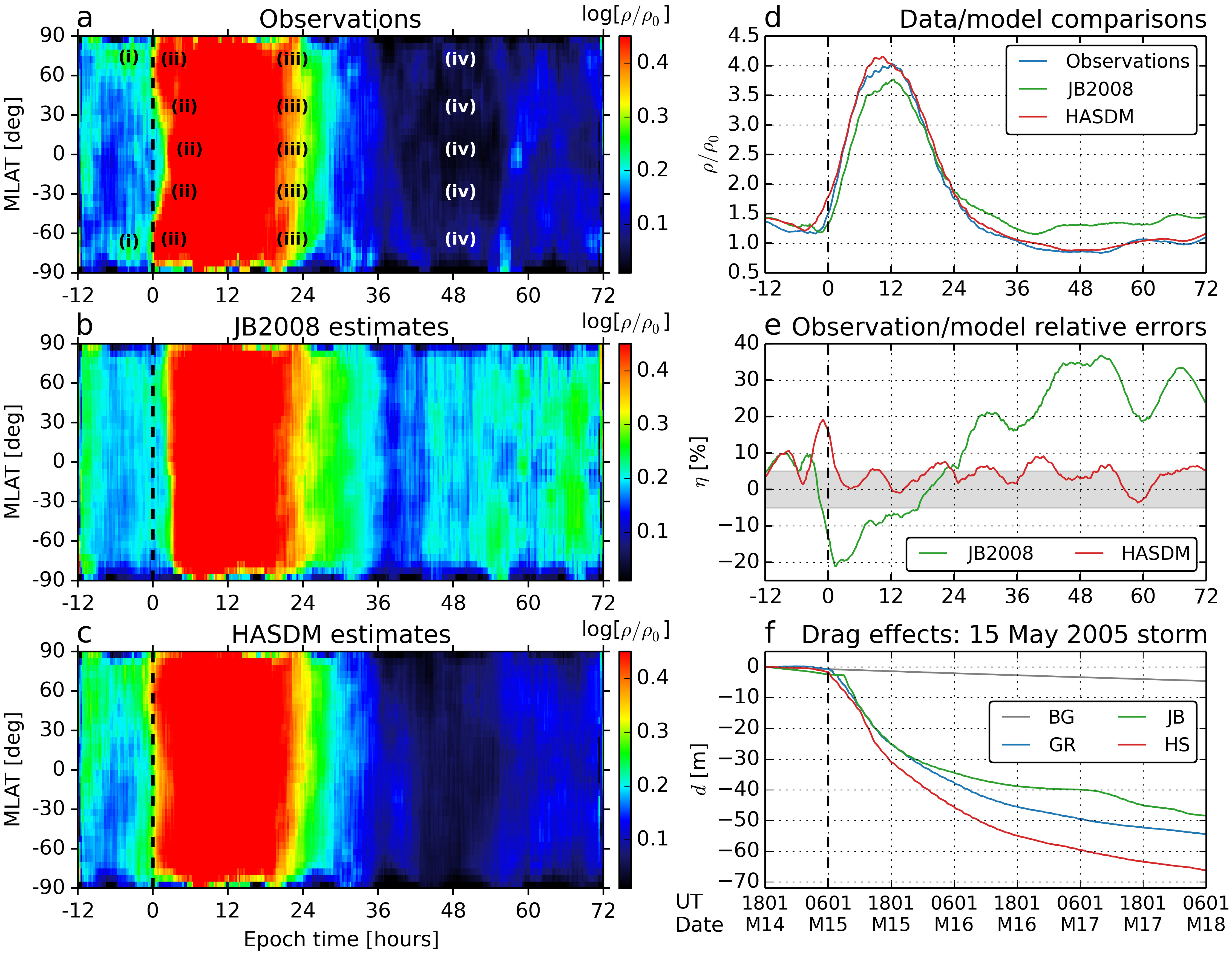}% This is a *.eps file
			\caption{Left column: superposed epoch analysis results of thermospheric neutral mass density response using the methodology introduced by \cite{Oliveira2017c} with observations (a), JB2008 estimates (b), and HASDM estimates (c). Right column: density averaged within --50$^\circ$ $\leq$ MLAT $\leq$ 50$^\circ$ in a 15-minute cadence (d); density estimate errors with respect to observations computed as $\eta_i$ = ($\rho_i$ - $\rho_{obs}$)/$\rho_{obs}$, with $i$ = (JB2008, HASDM) (e); storm-time drag effects computed for GRACE's orbit during the 15 May 2005 extreme storm using the framework provided by \cite{Oliveira2019b} (f). }
			\label{jbhs}
		\end{figure}

		The left column of Figure \ref{jbhs} shows thermosphere density response to the extreme storms whose coverages are shown in Figure \ref{xht}c and dates shown in Table 1 of \cite{Zesta2019a}. Density data are superposed in epoch time $\times$ MLAT bins with size 15 minutes $\times$ 3$^\circ$. The zero epoch time is taken as the time of each respective storm main phase onset. The panels show $\log[\rho/\rho_0]$, with $\rho$ being the storm-time observed or modeled density and $\rho_0$ being the background density estimated by JB2008 if there was no storm. \par

		As described by \cite{Oliveira2017c}, $\rho_0$ is obtained by the following steps: (i) JB2008 is used to compute the neutral density $\rho_1$ during periods of low geomagnetic activity ($|$Dst$|$ $<$ 30 nT) to exclude effects of high magnetospheric compressions and intense magnetic storms; (ii) a polynomial expansion \citep{Arlinghaus1994} with degree 15 is fitted to the ratio $\rho/\rho_1$ to obtain a calibration function $f(t)$ (this degree order provided the best polynomial fitting for a few control events, including moderate and intense storms); and (iii) with density computed by JB2008 with Dst = 0, i.e., $\rho(Dst = 0)$, and f(t) interpolated for the whole dataset, the background density $\rho_0$ is given by:
			\begin{equation}
				\rho_0 = \rho(Dst = 0)\times f(t)
			\end{equation}

		Results shown in Figure \ref{jbhs}a were published by \cite{Zesta2019a}, but here we removed effects caused by double storms, i.e., events in periods when another CME impacted Earth when the magnetosphere was highly driven. The data clearly show (i) high-latitude density enhancements due to the impacts of shocks in the driver leading edges; (ii) TAD propagation effects from high to low latitudes after storm main phase onset within 1-2 orbits ($\sim$2 hours, (iii) thermosphere cooling effects at all latitudes $\sim$22 hours during storm time presumably due to Nitric Oxide (NO) cooling effects \citep{Mlynczak2003,Knipp2017a,Zesta2019a} and (iv) thermosphere overcooling during storm recovery meaning that density levels are now lower than density levels during pre-storm periods \citep{Lei2012,Zhang2019}. See Roman numerals in panel a for the corresponding effect described above on density data. \par

		The comparison between JB2008 estimates (b) with observations (a) is quite remarkable. JB2008 does not capture effects caused by pre-storm shock compressions, nor does it capture TAD effects at storm main phase onset (white arrows in panel a). Additionally, JB2008 predicts thermosphere heating at low and high latitudes at the same time, which coincides with the time TADs take to reach equatorial latitudes as shown by observations. Although JB2008 predicts a major thermosphere cooling at $\sim$22 hours, the model predicts a secondary cooling occurring $\sim$8 hours later in comparison to observations, and the model does not predict thermosphere overcooling effects during the storm recovery phase. Quite the contrary, densities are slightly higher during recovery in comparison to pre-storm density levels. \par

		HASDM results (c) are considerably improved in comparison to JB2008 results. Although HASDM does not clearly show high-latitude thermosphere heating due to pre-storm shock compressions, the model captures some patterns of TAD propagation, the extreme latitudinal cooling at t $\sim$ 22 hours, and the thermosphere overcooling during storm recovery. Panel d shows mid- and low-latitude densities averaged within -- 50$^\circ$ $\leq$ MLAT $\leq$ 50$^\circ$ every 15 minutes throughout the storms. Our results show that (i) JB2008 and HASDM densities agree moderately well with observations during pre-storm periods; (ii) JB2008 underestimates and HASDM overestimates densities during storm main phase, and (iii) HASDM estimates agree remarkably well with observations during storm recovery, reproducing cooling effects quite well, but JB2008 density levels always surpass observations during the recovery phase. \cite{Licata2021b} noted similar trends when comparing CHAMP and GRACE observations with HASDM and JB2008 results for the Halloween storms. Panel e shows the relative errors of the model estimates with respect to observations. The shaded grey area corresponds to the $\pm$5\% confidence interval suggested by the U.S. Space Force for optimizing orbit predictions \citep{Lewis2019}. Results show that HASDM errors stay within the confidence interval during all storm periods, but the model introduces high errors during times of pre-storm shock compressions. On the other hand, the standalone JB2008 performance is considerably worse during all storm periods. The model underestimates density during the main phase, but it considerably overestimates density during the recovery phase, reaching error levels as high as 35\% during late recovery. As discussed in \cite{Oliveira2019b} and \cite{Zesta2019a}, JB2008 should be modified to accommodate effects induced by NO cooling for the improvement of density predictions particularly during extreme magnetic storms. \par

		Finally, Figure \ref{jbhs}f shows the comparison of observed and modeled drag effects on GRACE during the 15 May 2005 extreme magnetic storm. The drag effects are computed according to the framework developed by \cite{Oliveira2019b}. The nearly straight grey line corresponds to the drag effect caused by the background density $\rho_0$ if there was no storm. Observations (blue curve) show that GRACE decayed by 55 m 72 hours after storm onset. This is 11 times more severe in comparison to the orbital decay caused by the background density. As a result of the errors introduced by JB2008 and HASDM during different phases of the storm cycle (e), drag effects are underestimated by the former and overestimated by the latter.

	\section{Discussion and conclusion}

		The Sputnik satellites opened the doors for human exploration in space \citep{Launius2004}. \cite{Jacchia1959} used ephemeris data collected by Sputnik during an extreme magnetic storm in 1958 to link orbital drag effects with geomagnetic activity for the first time. Many empirical models were created and satellite missions were designed to study the dynamic thermosphere response to magnetic storms throughout the decades. However, orbital drag effects induced in LEO by extreme storms are yet relatively less understood partly due to the rarity of extreme storms during the space age. The International Geophysical Year (IGY) is marked by the adoption of the Dst index that became a standard trademark in assessing the intensity of magnetic storms. Interestingly, the largest yearly sunspot number ever recorded occurred in 1957 \citep[e.g.,][]{Clette2016b}, only a year before Sputnik launched and observed intense orbital drag effects \citep{Jacchia1959}. Although human exploration in space had just begun, the overall solar activity throughout the solar cycles (SCs) has been steadily decreasing since then. Since the IGY, relatively few extreme events (Dst $\leq$ -- 250 nT) have occurred. \cite{Meng2019} reported on the occurrence of $\sim40$ extreme events, from which only 7 were observed by spacecraft with high-precision accelerometers for density derivations. The last extreme magnetic storm took place two SCs ago (SC23) on 15 May 2005. \par

		Given the very low number of extreme magnetic storms amongst hundreds of events on record, a way to investigate more extreme events is to look in the past. Recently, many efforts have been undertaken to investigate magnetic activity represented by Dst-like indices reconstructed from archival materials generally drafted in the first half of the 20th century. Some of these events occurred in October/November 1903 \citep{Hayakawa2020a}, September 1909 \citep{Love2019a,Hayakawa2019a}, May 1921 \citep{Love2019b}, and March 1946 \citep{Hayakawa2020b}. These works showed that minimum Dst-like values of these storms went below -- 500 nT, which characterizes the event as a superstorm. Each reference provides the respective Dst-like index data for further investigations. For example, \cite{Oliveira2020b} used the Dst-like data for the first 3 events mentioned above along with real Dst data for the March 1989 storm \citep{Allen1989,Boteler2019} to show with JB2008 that, by comparison, long-lasting and less intense superstorms can induce more severe drag effects than short-lasting and more intense superstorms. Further investigations are needed here, as we have yet the unexplored events occurring in September 1859 \citep{Tsurutani2003b,Hayakawa2019b}, February 1872 \citep{Silverman2008,Hayakawa2018a}, and March 1946 \citep{Hayakawa2020b}. \par

		Now looking to the future, a possibility to improve our understanding of extreme orbital drag is to foster international collaboration for the development of a central database of thermospheric density data. There has been increased interest of the private sector in the elaboration and construction of a very large constellation of satellites in LEO. For example, SpaceX launched in 2018 the first satellite prototypes for the Starlink project \citep{SpaceX2016}. Starlink's primary goal is to create a megaconstellation of over 12,000 LEO satellites bellow 600 km altitude with latitudinal distribution of 0.005-0.01 spacecraft per square meter for the formation of a worldwide internet network \citep{McDowell2020}. Another private company, OneWeb, intends so launch by the end of 2022 a 648-satellite constellation to provide worldwide internet service as well \citep{Barnett2016}. Using an evolutionary model with parameters provided by the respective operations application \citep{Barnett2016,SpaceX2016}, \cite{Lemay2018} showed that, within a 5-year operation time, the probability of occurring a catastrophic collision involving an OneWeb spacecraft is $\sim$5\%, whereas the same for SpaceX is near 50\%. However, the authors did not consider any effects introduced by solar activity, and, if they did, these figures would have certainly been higher. Therefore, the need of accurate models for orbital track prediction during magnetic storms, particularly during extreme events, is an important tool for preventing the occurrence of the Kessler Syndrome in space. \par

		A future large-scale thermospheric density data base can be used in studies involving Machine Learning (ML) applications. ML studies have become very popular in the field of Earth and Space Sciences in the past decade \citep[e.g., ][]{Keesee2020,Smith2020a,Bortnik2021,Haines2021}. For example, historical data sets (geomagnetic and solar indices, sunspot numbers) prior to 2000 can be used for training a model to predict storm drivers and the subsequent global thermospheric density and orbital drag of a LEO satellite in a given location \citep{Licata2020,Licata2021c}. This can be accomplished, e.g., by the use of ML linear regression for ``training" the model to ``learn" how to predict these effects \citep{Rong2018}. Such technique is named by \cite{Bortnik2021} as {\it Time Series and Spatiotemporal Prediction}, and is recognized by the authors as an important step for ML applications in Earth and Space Sciences. As shown in Figure \ref{jbhs}, if HASDM provided remarkable results with the use of only 75 calibrated satellites, such results can be further improved by the assimilation of large-scale data provided by a few hundreds satellites in LEO, particularly for extreme events. \par

		Finally, recent studies suggest that solar activity of SC25 will be approximately the same as the solar activity of SC24 \citep[e.g.,][]{Javaraiah2017}. However, \cite{McIntosh2020} predict that SC25 will not only be stronger than SC24, but SC25 will rival the magnitude of the strongest SCs on record. Additionally, \cite{Javaraiah2017} suggests that the transition between SC25 and SC26 will coincide with the end of the current Gleissberg cycle. The Gleisseberg cycle is characterized by the periodic occurrences of solar maxima every 77-88 years or so \citep{Gleissberg1967,Feynman2014}. Therefore, all these predictions indicate that in the next years and decades the Sun will shift from its relatively quiet conditions to a much more active behavior. As a result, very accurate and precise thermospheric density models will play crucial roles in keeping and guaranteeing the safety of the ever increasing number of satellites in LEO by improved orbital tracking particularly during extreme magnetic storms.

	\section*{Conflict of Interest Statement}
	
		The authors declare that the research was conducted in the absence of any commercial or financial relationships that could be construed as a potential conflict of interest.

	\section*{Author Contributions}

		DMO is the primary author of the article and performed the superposed epoch analyses and drag computations. EZ contributed with the overall discussions and helped the primary author set up the overall structure of the manuscript. PMM and RLL performed the computations of the JB2008 and SET HASDM densities for the extreme events. WKT and MDP provided support for the SET HASDM density computations. HH provided importance guidance with respect to historical data set discussions and interpretation of past and future solar activity discussions. All co-authors provided editorial comments thus contributing to the article and approved the submitted version.

	\section*{Funding}

		This work was funded by the NASA Space Weather Science Applications Operations 2 Research program, under grant number 20-SWO2R20-2-0014.

	\section*{Data Availability Statement}

		LEO satellite (ephemeris and density) data for CHAMP, GRACE, GOCE, and Swarm were obtained from the repository provided by the Delft University of Technology (\url{http://thermosphere.tudelft.nl}). Although not explicitly shown, geomagnetic index data represented by SYM-H (1-minute resolution), were downloaded from the \cite{WDC_AE2015}. The JB2008 model is publicly available at the Space Environment Technologies (SET) website (\url{https://spacewx.com/jb2008/}). Dst and solar index data are also provided by SET with the source codes. The JB2008 and SET HASDM outputs can be obtained from the repository located at \url{https://zenodo.org/record/4602380#.YRwSXC2cYkh}.

	\section*{Acknowledgments}

		The authors of this manuscript warmly thank Delft University of Technology for providing and making public the LEO satellite data used in this study. HH thanks World Data Center for the production, preservation and dissemination of the international sunspot number (WDC SILSO) and Data Analysis Center for Geomagnetism and Space Magnetism (WDC for Geomagnetism at Kyoto) for providing the international sunspot number and the Dst index.

	\bibliographystyle{frontiersinSCNS_ENG_HUMS} 
	%\bibliography{Oliveira_main}

% Manuscript contribution to the field (required by Frontiers for submission)

% Thermosphere neutral mass density is strongly enhanced during magnetic storms, particularly during extreme events. Density is one of the most important factors for determining the subsequent orbital drag on satellites in the upper atmosphere. Orbital drag can reduce a satellite’s lifetime and introduce large errors and uncertainties as the storm becomes more intense. Given the planned increase of government and private satellite presence in LEO, the need for accurate density predictions for collision avoidance and lifetime optimization, particularly during extreme events, has become an urgent matter and requires comprehensive international collaboration. In this article, we summarize our knowledge of thermospheric neutral mass density response to extreme storms and compare model capabilities in predicting thermosphere heating during extreme events. We discuss how forecasting models can be improved by looking into two directions: first, to the past, by adapting historical extreme storm datasets for density predictions, and second, to the future, by facilitating the assimilation of large-scale thermosphere data sets that will be collected in future events. Therefore, this topic is relevant to the scientific community, the U.S. Federal Government, and the private sector with assets operating in LEO.

\end{document}